\documentstyle{article}
\begin{document}

\title{\flushright{\small KL-01/03} \bigskip \bigskip \bigskip \bigskip
\bigskip \bigskip \\
\center{Superconformal and super B.R.S. invariance
             of the  N=1 supersymmetric W. Z. W. model 
             based on  Lie superalgebra}\thanks{%
Supported by DAAD}}
\author{L. Mesref\thanks{%
Email: lmesref@physik.uni-kl.de}}
\date{Department of Physics, Theoretical Physics\\
University of Kaiserslautern, Postfach 3049\\
67653 Kaiserslautern, Germany}
\maketitle

\begin{abstract}
We \ study the superconformal and super B.R.S. invariance of \ 
supersymmetric Wess Zumino Witten model based on Lie superalgebra. The
computation of the critical super dimension of this model is done using the
Fujikawa regularisation. finally, we recover the well known result which
fixes the relative coupling constant $\alpha ^{2}=1$ in a rigorous way.
\newpage {}
\end{abstract}

\section{Introduction}

The Wess Zumino Witten model \cite{witten} in two dimensions plays a central
role in many current investigations. A tremendous amount of literature has
been devoted to this model \cite{diviccia}. In this paper, we study the
superconformal and super B.R.S. invariance of the supersymmetric extension of
this model based on Lie superalgebra using the Fujikawa regularization \cite
{fujikawa}. The regularization allows us to evaluate the anomalous
superconformal and super B.R.S. currents extending the results of Ref [4] to
 N=1 supersymmetric case. The expression of the super Jacobian exhibits forms
of anomalies which are in agreement with several results based essentially
on the computation of the supersymmetric function at one loop level \cite
{abdalla}. We recover the well known result which fixes the relative
coupling constant $\alpha ^{2}=1$ in a rigorous way.

This paper is organized as follows: In Sec 2, we present the Wess Zumino
Witten action. In Sec 3, we study its invariance under the superconformal
transformations and their corresponding super B.R.S transformations. In Sec
4, we compute the super Jacobian of the holomorphic super B.R.S
transformations of the path integral measure and we derive the super B.R.S
identity using the Fujikawa regularization.

\section{The Wess Zumino Witten action}

Let us first consider the action describing this model \cite{abdalla}. In
the covariant superconformal gauge \cite{martinec}, this action is given by:

\bigskip

\begin{eqnarray}
A_{U} &=&\int_{\Sigma _{2}}\frac{d^{2}zd^{2}\theta }{8\pi \alpha ^{\prime }}%
\frac{R^{2}}{2}Tr\left( \overline{D}UDU^{-1}\right)   \nonumber \\
&&+\frac{k}{N_{G}}\int_{\Sigma _{3}}dtd^{2}\xi d^{2}\theta Tr\,U^{-1}%
\stackrel{.}{U}\left( U^{-1}D^{\alpha }U\right) \,\left( -i\gamma
_{5}\right) _{\beta }^{\alpha }\left( U^{-1}D_{\beta }U\right) , 
\end{eqnarray}

\bigskip 

where U is an element of a generic compact Lie supergroup G associated to a
Lie superalgebra g, D and $\overline{D}$ are complex super derivatives, $%
\Sigma _{3}$ is a three super Riemann surface and $\Sigma _{2}$ its
boundary $\left( \partial \Sigma _{3}=\Sigma _{2}\right) $ and $N_{G}
$ is a normalization factor. $\alpha ^{\prime }$ is the Regge slope which is
related to the string tension by: $T=\frac{1}{2\pi \alpha ^{\prime }}$ and R
is the radius of the compact group supermanifold. It measures the size of
the supermanifold. Low energy states are associated with the propagation of
the particles on the Minkowski space time $M^{4}$ . Any propagation on G
will require energies $\sim hc/R.$ Such states would be unexcited by probes
of energy $\ll hc/R$ and the internal space would therefore be invisible.
Then if one assumes the motion space to be a direct product of Minkowski
space with an internal group space G, we may tentatively choose the
dimension of Minkowski space to $d=4$ by a suitable choice of the internal
group G. If this is done, the model turn out to be a superstring field
theory in four dimensional space-time. This little digression has a great
significance in the mechanism of compactification of string theory \cite
{martinec} and Kaluza-Klein theories \cite{appelquist}. A proper job would
require us to go into details that are not crucial to the present discussion.

Although the action specified in Eq (1) is multivalued, the factor $\exp
\left( -A_{U}\right) $ is unique, provided the coupling constant k in front
of the Wess Zumino term is an integer. The action (1) can be seen as the
action of a closed type II supersymmetric string (two-dimensional
supergravity) or supersymmetric non linear defined on a compact group
supermanifold built on Lie superalgebra in the presence of the Wess Zumino
term. This latter term was originally introduced by Witten \cite{witten} to
restore the conformal invariance at the quantum level. In the context of
non-abelian bosonization, Witten has shown that if the relation: $\frac{%
\alpha ^{\prime }}{R^{2}}=\frac{N_{G}}{24\left| k\right| }$ ,i.e. $\alpha
^{2}=1$ is satisfied, conformal invariance and hence reparametrization
invariance is restored. This relation means that the radius of the
compactified dimensions gets quantized in units of the string tension.
Furthermore, when k approaches infinity one recovers the flat space limit.
The Wess Zumino term is topological and can be also interpreted as a torsion
which parallelizes the curvative at the point which the $\beta $-function
vanishes \cite{abdalla}.

\section{Gauge fixing}

The covariant superconformal gauge choice yields to the action of the
Fadeev-Popov ghost superfield system (B,C ) of superconformal spin $\left(
3/2,-1\right) $

\bigskip

\begin{equation}
A_{gh}=\int \frac{d^{2}zd^{2}\theta }{4\pi \alpha ^{\prime }}\left( B%
\overline{D}C+\overline{B}D\overline{C}\right) . 
\end{equation}

\bigskip 

Under the superconformal transformations parametrized by infinitesimal super
vector fields V and $\overline{V}$ satisfying the conditions:

\bigskip

\begin{equation}
\overline{\partial }V=\overline{D}V=D\overline{V}=0
\end{equation}

\bigskip

the superfields variables of matter, the ghost and the superconformal factor
transform as \cite{friedan}:

\bigskip 

\begin{equation}
\delta U=V\partial U+\frac{1}{2}\left( DV\right) DU+\overline{V}\overline{%
\partial }U+\frac{1}{2}\left( \overline{D}\overline{V}\right) \overline{D}U,
\end{equation}

\bigskip 

\begin{equation}
\delta C=V\partial C+\frac{1}{2}\left( DV\right) DC-\left( \partial V\right)
C+\overline{V}\overline{\partial }C+\frac{1}{2}\left( \overline{D}\overline{V%
}\right) \overline{D}C, 
\end{equation}

\bigskip 

\begin{equation}
\delta \overline{C}=V\partial \overline{C}+\frac{1}{2}\left( DV\right) D%
\overline{C}-\left( \partial V\right) \overline{C}+\overline{V}\overline{%
\partial }\overline{C}+\frac{1}{2}\left( \overline{D}\overline{V}\right)
\overline{D}\overline{C},
\end{equation}

\bigskip

\begin{equation}
\delta B=V\partial B+\frac{1}{2}\left( DV\right) DB+\frac{3}{2}\left(
\partial V\right) B+\overline{V}\overline{\partial }B+\frac{1}{2}\left(
\overline{D}\overline{V}\right) \overline{D}B,
\end{equation}

\bigskip

\begin{equation}
\delta \overline{B}=V\partial \overline{B}+\frac{1}{2}\left( DV\right) D%
\overline{B}+\frac{3}{2}\left( \overline{\partial }\overline{V}\right)
\overline{B}+\overline{V}\overline{\partial }\overline{B}+\frac{1}{2}\left( 
\overline{D}\overline{V}\right) \overline{D}\overline{B},  
\end{equation}

\bigskip 

\begin{equation}
\delta \Psi =V\partial \Psi +\frac{1}{2}\left( DV\right) D\Psi -\frac{1}{4}%
\left( \partial V\right) +\overline{V}\overline{\partial }\Psi +\frac{1}{2}%
\left( \overline{D}\overline{V}\right) \overline{D}\Psi -\frac{1}{4}%
\overline{\partial }\overline{V}.
\end{equation}

\bigskip

Under these transformations, the action changes as:

\bigskip

\begin{equation}
\delta A_{U+gh}=\int \frac{d^{2}zd^{2}\theta }{4\pi \alpha ^{\prime }}\left[
V\overline{D}\left( V\overline{D}\left( T^{U}+T^{gh}\right) _{z\theta
}\right) +h.c.\right] , 
\end{equation}

\bigskip

where $T_{z\theta }^{U}+T_{z\theta }^{gh}$ is the super stress-energy tensor
given by:

\bigskip 

\begin{equation}
T_{z\theta }^{U}=-\frac{R^{2}}{2}Tr\left( \partial UDU^{-1}\right) ,
\end{equation}

\bigskip

\begin{equation}
T_{z\theta }^{gh}=-C\partial B+\frac{1}{2}\left( DC\right) DB-\frac{3}{2}%
\left( \partial C\right) B.  
\end{equation}

\bigskip 

These currents are separately conserved:

\bigskip 

\begin{equation}
\overline{D}T_{z\theta }^{U}=\overline{D}T_{z\theta }^{gh}=D\overline{T}%
_{z\theta }^{U}=D\overline{T}_{z\theta }^{gh}=0. 
\end{equation}

\bigskip

The action $A_{U+gh}$ is also invariant under the super-BRS transformations
obtained by the replacement, in the superconformal transformations of U and $%
\Psi $, of V and $\overline{V}$ by respectively $\epsilon C$ and $\overline{%
\epsilon }\overline{C}$ :

\bigskip 

\begin{equation}
\delta U=\epsilon C\partial U-\frac{1}{2}\left( DC\right) DU+\overline{%
\epsilon }\overline{C}\overline{\partial }U-\frac{1}{2}\overline{\epsilon }%
\left( \overline{D}\overline{C}\right) \overline{D}U,  
\end{equation}

\bigskip 

\begin{equation}
\delta \Psi =\epsilon C\partial \Psi -\frac{1}{2}\epsilon \left( DC\right)
D\Psi -\frac{1}{4}\epsilon \partial C+\overline{\epsilon }\overline{C}%
\overline{\partial }\Psi -\frac{1}{2}\overline{\epsilon }\left( \overline{D}%
\overline{C}\right) \overline{D}\Psi -\frac{1}{4}\overline{\epsilon }%
\overline{\partial }C,  
\end{equation}

\bigskip 

where $\epsilon $ is a Grassmann parameter. For the ghost superfield system,
the super-BRS transformations are defined as:

\bigskip

\begin{equation}
\delta C=\epsilon C\partial C-\frac{1}{4}\epsilon \left( DC\right) \left(
DC\right) +\overline{\epsilon }\overline{C}\overline{\partial }C-\frac{1}{4}%
\overline{\epsilon }\left( \overline{D}\overline{C}\right) \left( \overline{D%
}C\right) ,  
\end{equation}

\bigskip 

\begin{equation}
\delta \overline{C}=\epsilon C\partial \overline{C}-\frac{1}{4}\epsilon
\left( DC\right) \left( D\overline{C}\right) +\overline{\epsilon }\overline{C%
}\overline{\partial }\overline{C}-\frac{1}{4}\overline{\epsilon }\left(
\overline{D}\overline{C}\right) \left( \overline{D}\overline{C}\right) , 
\end{equation}

\bigskip 

\begin{equation}
\delta B=\epsilon \left( T^{U}+T^{gh}\right) +\overline{\epsilon }\overline{C%
}\overline{\partial }B-\frac{1}{2}\overline{\epsilon }\left( \overline{D}%
\overline{C}\right) \overline{D}B,  
\end{equation}

\bigskip 

\begin{equation}
\delta \overline{B}=\overline{\epsilon }\left( \overline{T}^{U}+\overline{T}%
^{gh}\right) +\epsilon C\partial \overline{B}-\frac{1}{2}\epsilon \left(
DC\right) D\overline{B}. 
\end{equation}

\bigskip 

All these invariances are verified in the holomorphic sector where we
consider the variations, with $\overline{V}=\overline{\epsilon }=0$, of the
spinning string variable and its ghost only with the superconformal factor $%
\Psi $ and the anti-ghost superfield system kept fixed.

Under the localised holomorphic super B.R.S transformations which we
will consider in the next section,

\bigskip 

\begin{equation}
\delta U=\epsilon \left( x\right) C\partial U-\frac{1}{2}\epsilon \left(
x\right) \left( DC\right) \left( DU\right) , 
\end{equation}

\bigskip

\begin{equation}
\delta C=\epsilon \left( x\right) C\partial C-\frac{1}{4}\epsilon \left(
x\right) \left( DC\right) \left( DC\right) ,  
\end{equation}

\bigskip 

\begin{equation}
\delta B=\epsilon \left( x\right) \left( T^{U}+T^{gh}\right) ,
\end{equation}

\bigskip 

\begin{equation}
\delta \overline{B}=\delta \overline{C}=\delta \Psi =0,  
\end{equation}

\bigskip 

the action changes as 

\bigskip 

\begin{eqnarray}
\delta A_{U+gh} &=&\int \frac{d^{2}zd^{2}\theta }{4\pi \alpha ^{\prime }}%
\left[ \left( \overline{D}\epsilon \right) J_{\theta }+\frac{1}{4}\left( 
\overline{D}\epsilon \right) D\left( BCDC\right) \right] +\int \frac{%
d^{2}zd^{2}\theta }{4\pi \alpha ^{\prime }}\left( D\epsilon \right)  
\nonumber \\
&&\times \left[ -C\left[ \overline{D}DUDU^{-1}\right] -D\left( CB\overline{D}%
C\right) +\frac{1}{2}BDCD\overline{C}\right] , 
\end{eqnarray}

\bigskip 

with the canonical super B.R.S current

\bigskip 

\begin{equation}
J_{\theta }=C\left( T_{z\theta }^{U}+\frac{1}{2}T_{z\theta }^{gh}\right) =%
\frac{-R^{2}}{2}C\partial UDU^{-1}-C\left[ \frac{3}{4}BDC-\frac{1}{4}\left(
DB\right) DC\right] .  
\end{equation}

\bigskip 

This current is conserved when we use the classical equations of motion for
the superfields U, B and C.

\qquad Before we close this section, we make one remark concerning the
superfield U. In the quantum theory, the superfield fluctuates around the
classical solution, and to analyse these fluctuations, we set:

\bigskip

\begin{equation}
U\left( z,\overline{z},\theta ,\overline{\theta }\right) =U_{cl}\left( z,%
\overline{z},\theta ,\overline{\theta }\right) U_{q}\left( z,\overline{z}%
,\theta ,\overline{\theta }\right)  
\end{equation}

\bigskip

so that $U_{q}\left( z,\overline{z},\theta ,\overline{\theta }\right) $
fluctuates around $U_{q}\left( z,\overline{z},\theta ,\overline{\theta }%
\right) =1.$ We treat the amplitude of the space-independent fluctuations as
a collective variable, setting $U_{q}\left( z,\overline{z},\theta ,\overline{%
\theta }\right) =u_{0}\cdot \exp \left( X\left( z,\overline{z},\theta ,%
\overline{\theta }\right) +Y\left( z,\overline{z},\theta ,\overline{\theta }%
\right) \right) .$ Here is $u_{0}$ the collective variable associated with
zero modes, while $X+Y$ describes the fluctuations in the nonzero modes. The
superfied U is then given by :

\bigskip

\begin{equation}
U=U_{cl}\cdot u_{0}\cdot \exp \left( X+Y\right) =U_{0}\cdot \exp \left(
X+Y\right) =U_{0}\cdot \exp \left( X^{A}T_{A}+Y^{a}S_{a}\right) ,  
\end{equation}

\bigskip 

where the bosonic generators,

\bigskip 

\begin{equation}
T_{A},\ A=1,...,D_{G}  
\end{equation}

and the fermionic generators

\bigskip 

\begin{equation}
S_{a},\ a=1....,d_{G}
\end{equation}

\bigskip 

span the superalgebra g associated to a generic Lie supergroup G. The Lie
product

\bigskip

\begin{equation}
\left[ T_{A},T_{B}\right] =f_{AB}^{C}T_{C},  
\end{equation}

\bigskip

\begin{equation}
\left[ S_{a},S_{b}\right] _{+}=g_{ab}^{A}T_{A},  
\end{equation}

\bigskip

\begin{equation}
\left[ S_{a},T_{A}\right] =-\left[ T_{A},S_{a}\right] =h_{aA}^{b}S_{b} 
\end{equation}

\bigskip 

fulfills the Jacobi identities

\bigskip 

\begin{equation}
f_{AB}^{D}f_{CD}^{E}+f_{BC}^{D}f_{AD}^{E}+f_{CA}^{D}f_{BD}^{E}=0,
\end{equation}

\bigskip

\begin{equation}
h_{aC}^{b}f_{AB}^{C}+h_{aB}^{c}h_{cA}^{b}-h_{aA}^{c}h_{cB}^{b}=0, 
\end{equation}

\bigskip 

\begin{equation}
h_{bA}^{c}g_{ac}^{B}+h_{aA}^{c}g_{bc}^{B}+g_{ab}^{C}f_{AC}^{B}=0,
\end{equation}

\bigskip

\begin{equation}
g_{bc}^{A}h_{aA}^{d}+g_{ca}^{A}h_{bA}^{d}+g_{ab}^{A}h_{cA}^{d}=0. 
\end{equation}

\bigskip 

The generators $T_{A}$ and $S_{a}$\ are normalized as:

\bigskip 

\begin{equation}
Tr\left( T_{A}T_{B}\right) =2\delta _{AB},  
\end{equation}

\bigskip 

\begin{equation}
Tr\left( S_{a}S_{b}\right) =2\varepsilon _{ab},  
\end{equation}

\bigskip 

\begin{equation}
Tr\left( T_{A}S_{a}\right) =Tr\left( S_{a}T_{A}\right) =0.  
\end{equation}

\bigskip 

Notice that $X^{A}T_{A}=T_{A}X^{A}$ since they are bosons and $%
Y^{a}S_{a}=-S_{a}Y^{a}$ since they are fermions. The superfields and
transform under the superconformal transformations as:

\bigskip 

\begin{equation}
\delta X^{A}=V\partial X^{A}+\frac{1}{2}\left( DV\right) DX^{A}+\overline{V}%
\overline{\partial }X^{A}+\frac{1}{2}\left( \overline{D}\overline{V}\right) 
\overline{D}X^{A},  
\end{equation}

\bigskip 

\begin{equation}
\delta Y^{a}=V\partial Y^{a}+\frac{1}{2}\left( DV\right) DY^{a}+\overline{V}%
\overline{\partial }Y^{a}+\frac{1}{2}\left( \overline{D}\overline{V}\right) 
\overline{D}Y^{a} 
\end{equation}

\bigskip 

and transform under the super B.R.S transformations as:

\bigskip 

\begin{equation}
\delta X^{A}=\epsilon C\partial X^{A}-\frac{1}{2}\epsilon \left( DC\right)
DX^{A}+\overline{\epsilon }\overline{C}\overline{\partial }X^{A}-\frac{1}{2}%
\overline{\epsilon }\left( \overline{D}\overline{C}\right) \overline{D}X^{A},
\end{equation}

\bigskip

\begin{equation}
\delta Y^{a}=\epsilon C\partial Y^{a}-\frac{1}{2}\epsilon \left( DC\right)
DY^{a}+\overline{\epsilon }\overline{C}\overline{\partial }Y^{a}-\frac{1}{2}%
\overline{\epsilon }\left( \overline{D}\overline{C}\right) \overline{D}Y^{a}.
\end{equation}

\bigskip 

In terms of the $X^{A}$ and $Y^{a}$, the action becomes:

\bigskip 

\begin{eqnarray}
A_{U}\left( U_{0},X^{A},Y^{a}\right)  &=&A\left( U_{0}\right) +\int \frac{%
d^{2}zd^{2}\theta }{4\pi \alpha 
%TCIMACRO{\UNICODE[m]{0xb4}}%
%BeginExpansion
{\acute{}}%
%EndExpansion
}R^{2}\left[ -\overline{D}X^{A}DX^{B}\delta _{AB}+\overline{D}%
Y^{a}DY^{b}\epsilon _{ab}\right]   \nonumber \\
&&+\int \frac{d^{2}zd^{2}\theta }{8\pi \alpha
%TCIMACRO{\UNICODE[m]{0xb4}}%
%BeginExpansion
{\acute{}}%
%EndExpansion
}R^{2}{\LARGE [}\frac{f_{AB}^{C}}{2}[Tr\left( U_{0}^{-1}\overline{D}%
U_{0}T_{C}\right) X^{A}DX^{B}\left( 1-\alpha \right)   \nonumber \\
&&+Tr\left( U_{0}^{-1}DU_{0}T_{C}\right) X^{A}\overline{D}X^{B}\left(
1+\alpha \right) ]  \nonumber \\
&&-\frac{g_{ab}^{A}}{2}[Tr\left( U_{0}^{-1}\overline{D}U_{0}T_{A}\right)
Y^{a}DY^{b}\left( 1-\alpha \right)   \nonumber \\
&&+Tr\left( U_{0}^{-1}DU_{0}T_{A}\right) Y^{a}\overline{D}Y^{b}\left(
1+\alpha \right) ]  \nonumber \\
&&+\frac{h_{aA}^{b}}{2}[Tr\left( U_{0}^{-1}\overline{D}U_{0}S_{b}\right)
X^{A}DY^{a}\left( 1-\alpha \right)   \nonumber \\
&&+Tr\left( U_{0}^{-1}DU_{0}S_{b}\right) X^{A}\overline{D}Y^{a}\left(
1+\alpha \right) ]  \nonumber \\
&&-\frac{h_{aA}^{b}}{2}[Tr\left( U_{0}^{-1}\overline{D}U_{0}S_{b}\right)
Y^{a}DX^{A}\left( 1-\alpha \right)   \nonumber \\
&&+Tr\left( U_{0}^{-1}DU_{0}S_{b}\right) Y^{a}\overline{D}X^{A}\left(
1+\alpha \right) ]{\LARGE ]},  
\end{eqnarray}

\bigskip 

where

\bigskip 

\begin{eqnarray}
A\left( U_{0}\right)  &=&\int_{\Sigma _{2}}\frac{d^{2}zd^{2}\theta }{8\pi
\alpha 
%TCIMACRO{\UNICODE[m]{0xb4}}%
%BeginExpansion
{\acute{}}%
%EndExpansion
}\frac{R^{2}}{2}\left( \overline{D}U_{0}DU_{0}^{-1}\right)    \\
&&+\frac{k}{N_{G}}\int_{\Sigma _{3}}dtd^{2}\xi d^{2}\theta TrU_{0}^{-1}\dot{%
U}_{0}\left( U_{0}^{-1}D^{\alpha }U_{0}\right) \left( -i\gamma _{5}\right)
_{\alpha }^{\beta }\left( U_{0}^{-1}D_{\beta }U_{0}\right) ,  \nonumber  
\end{eqnarray}

\bigskip

which gives:

\bigskip

\begin{equation}
A_{U}\left( U_{0},X^{A},Y^{a}\right) =\int \frac{d^{2}zd^{2}\theta }{4\pi
\alpha
%TCIMACRO{\UNICODE[m]{0xb4}}%
%BeginExpansion
{\acute{}}%
%EndExpansion
}L\left( U_{0},X^{A},Y^{a}\right)  
\end{equation}

\bigskip

up to terms cubic or higher order in $X^{A}$ and\ $Y^{a}$. We restricted our
selves to the quadratic order and used the classical equations of motion to
eliminate the first-order terms.

\bigskip

\section{Generating functional and super B.R.S identity}

Following, Polyakov \cite{polyakov} the partition function of the model at
hand is given in the superconformal gauge by:

\bigskip

\begin{equation}
Z=\int D_{\mu }e^{-A(\widetilde{X}^{A},\widetilde{Y}^{a},\widetilde{B},%
\widetilde{C})},  
\end{equation}

\bigskip 

where $D_{\mu }$ is the reparametrization invariant measure and we have
defined the new superfields as:

\bigskip

\begin{equation}
\widetilde{X}^{A}=e^{-\Psi }X^{A},  
\end{equation}

\bigskip

\begin{equation}
\widetilde{Y}^{a}=e^{-\Psi }Y^{a},  
\end{equation}

\bigskip

\begin{equation}
\widetilde{C}=e^{-3\Psi }C,  
\end{equation}

\bigskip

\begin{equation}
\overline{\widetilde{C}}=e^{-3\Psi }\overline{C},  
\end{equation}

\bigskip

\begin{equation}
\widetilde{B}=e^{2\Psi }B,  
\end{equation}

\bigskip

\begin{equation}
\overline{\widetilde{B}}=e^{2\Psi }\overline{B}. 
\end{equation}

\bigskip

This method is completely analogous to that followed by Fujikawa \cite
{fujikawa}. One may start with variables with suitable weight factors so the
resulting path integral measure is formally invariant under diffeomorphisms.

The way to derive super B.R.S identity, which is insensitive to the
definition of the path integral variable is to start with the localized
variations:

\bigskip

\begin{equation}
\delta \widetilde{X}^{A}=\epsilon \left( x\right) \left[ C\partial 
\widetilde{X}^{A}-\frac{1}{2}\left( DC\right) D\widetilde{X}^{A}+C\left(
\partial \Psi \right) \widetilde{X}^{A}-\frac{1}{2}\left( DC\right) \left(
D\Psi \right) \widetilde{X}^{A}\right] , 
\end{equation}

\bigskip 

\begin{equation}
\delta \widetilde{Y}^{a}=\epsilon \left( x\right) \left[ C\partial
\widetilde{Y}^{a}-\frac{1}{2}\left( DC\right) D\widetilde{Y}^{a}+C\left(
\partial \Psi \right) \widetilde{Y}^{a}-\frac{1}{2}\left( DC\right) \left(
D\Psi \right) \widetilde{Y}^{a}\right] ,  
\end{equation}

\bigskip

\begin{eqnarray}
\delta \widetilde{B} &=&\epsilon \left( x\right) {\LARGE [}C\partial 
\widetilde{B}-\frac{1}{2}\left( DC\right) D\widetilde{B}-2C\left( \partial
\Psi \right) \widetilde{B}+\left( DC\right) \left( D\Psi \right) \widetilde{B%
}  \nonumber \\
&&+\frac{3}{2}\left( \partial C\right) \widetilde{B}-e^{2\Psi }T_{z\theta
}^{X+Y}{\LARGE ]}, 
\end{eqnarray}

\bigskip 

\begin{equation}
\delta \widetilde{C}=\epsilon \left( x\right) \left[ C\partial \widetilde{C}-%
\frac{3}{4}\left( DC\right) \left( D\Psi \right) \widetilde{C}-\frac{1}{4}%
\left( DC\right) D\widetilde{C}\right] ,  
\end{equation}

\bigskip 

\begin{equation}
\delta \overline{\widetilde{B}}=\delta \overline{\widetilde{C}}=\delta \Psi
=0  
\end{equation}

To evaluate the super Jacobian factor resulting from these transformations
we use the differential of

\bigskip

\begin{eqnarray}
d\delta \widetilde{C} &=&\epsilon \left( x\right) {\LARGE [}C\partial 
\widetilde{C}-\frac{1}{2}\left( DC\right) D\left( d\widetilde{C}\right)
+3C\left( \partial \Psi \right) d\widetilde{C}  \nonumber \\
&&-\frac{3}{2}\left( DC\right) \left( D\Psi \right) d\widetilde{C}-\left(
\partial C\right) D\widetilde{C}{\LARGE ]}  
\end{eqnarray}

to obtain:

\bigskip

\begin{eqnarray}
\ln J &=&Str\ \epsilon \left( x\right) \left[ C\partial -\frac{1}{2}\left(
DC\right) D+C\left( \partial \Psi \right) -\frac{1}{2}\left( DC\right)
\left( D\Psi \right) \right] _{X^{A}}  \nonumber \\
&&+Str\ \epsilon \left( x\right) \left[ C\partial -\frac{1}{2}\left(
DC\right) D+C\left( \partial \Psi \right) -\frac{1}{2}\left( DC\right)
\left( D\Psi \right) \right] _{Y^{a}}  \nonumber \\
&&+Str\ \epsilon \left( x\right) \left[ C\partial -\frac{1}{2}\left(
DC\right) D+2C\left( \partial \Psi \right) +\left( DC\right) \left( D\Psi
\right) +\frac{3}{2}\left( \partial C\right) \right] _{B}  \nonumber \\
&&-Str\ \epsilon \left( x\right) \left[ C\partial -\frac{1}{2}\left(
DC\right) D+3C\left( \partial \Psi \right) -\frac{3}{2}\left( DC\right)
\left( D\Psi \right) -\left( \partial C\right) \right] _{C}.  
\end{eqnarray}

\bigskip 

The super-Jacobian factor of the ghost superfields, which are free, have
been evaluated in Ref. 3. For the Jacobian of the superfields $X^{A}$ and $%
Y^{a}$ we must regularize the ill-defined supertrace of the unity, $%
Str\left( 1\right) =\sum_{n=1}^{\infty }\sum_{A=1}^{D_{G}}\widetilde{\phi }%
_{n}^{A}\widetilde{\phi }_{n}^{A}$ and $Str\left( 1\right)
=\sum_{n=1}^{\infty }\sum_{a=1}^{d_{G}}\widetilde{\phi }_{n}^{a}\widetilde{%
\phi }_{n}^{a}.$ $\widetilde{\phi }_{n}^{A}$ and $\widetilde{\phi }_{n}^{a}$
belongs to a closed set of eigenfunctions of the Hermitian regulators:

\bigskip

\begin{equation}
H_{AB}\widetilde{\phi }_{n}^{B}=\lambda _{n}^{2}.\delta _{AB}.\widetilde{%
\phi }_{n}^{B}, 
\end{equation}

\bigskip 

\begin{equation}
H_{ab}\widetilde{\phi }_{n}^{b}=\lambda _{n}^{2}.\delta _{ab}.\widetilde{%
\phi }_{n}^{b},  
\end{equation}
where

\bigskip 

\begin{equation}
H_{AB}\widetilde{\phi }_{n}^{B}=\frac{\delta L}{\delta \widetilde{\phi }%
_{n}^{A}},  
\end{equation}

\bigskip

\begin{equation}
H_{ab}\widetilde{\phi }_{n}^{b}=\frac{\delta L}{\delta \widetilde{\phi }%
_{n}^{a}},  
\end{equation}

\bigskip 

L is defined in (46). The regulators $H_{AB}$ and $H_{ab}$ are
given by:

\bigskip

\begin{eqnarray}
H_{AB} &=&e^{\psi }\overline{D}D\delta _{AB}e^{\psi }-\frac{f_{ABC}}{4}%
[Tr\left( U_{0}^{-1}\overline{D}U_{0}T^{C}\right) e^{\psi }De^{\psi }\left(
1-\alpha \right)   \nonumber \\
&&+Tr\left( U_{0}^{-1}DU_{0}T^{C}\right) e^{\psi }\overline{D}e^{\psi
}\left( 1+\alpha \right) ],  
\end{eqnarray}

\bigskip

\begin{eqnarray}
H_{ab} &=&e^{\psi }\overline{D}D\delta _{ab}e^{\psi }-\frac{g_{abc}}{4}%
[Tr\left( U_{0}^{-1}\overline{D}U_{0}T^{C}\right) e^{\psi }De^{\psi }\left(
1-\alpha \right)   \nonumber \\
&&+Tr\left( U_{0}^{-1}DU_{0}T^{C}\right) e^{\psi }\overline{D}e^{\psi
}\left( 1+\alpha \right) ].  
\end{eqnarray}

\bigskip

There is no conceptual difficulty in the formulation of the regularization
since the regulators do not depend on the quantum fluctuations $X^{A}$ and \ 
$Y^{a}$. This is a result of the expansion in (44) which gives us a
quadratic action.

The evaluation of the ill defined supertrace in (60) is standard. One may
uses the basis of the super plane waves \ $e^{\left( ik\overline{z}+i%
\overline{k}z\right) -\chi \overline{\theta }-\overline{\chi }\theta }$ ,
introduces a cut-off M, rescales the variables $k\rightarrow Mk$ ,$\overline{k}$ $\rightarrow M\overline{k}$ , $\chi \rightarrow M\chi $ , $%
\overline{\chi }\rightarrow M\overline{\chi }$ and the fact that the
integration measure is invariant under this simultaneous rescaling of
bosonic and fermionic variables \cite{fujikawa}. This straightforward but
tedious calculation leads to:

\bigskip 

\begin{eqnarray}
\ln J &=&\frac{1}{2\pi }\int d^{2}zd^{2}\theta \epsilon \left( x\right)
{\LARGE [}5-\left( D_{G}-d_{G}\right) [C\left( \partial D\overline{D}\psi
\right) +2C\left( \partial \psi \right) \left( D\overline{D}\psi \right) ]
\nonumber \\
&&+10C\left( D\psi \right) \left( \partial \overline{D}\psi \right) +\frac{1%
}{2}\left( D_{G}-d_{G}\right) \left( DC\right) \left( \partial \overline{D}%
\psi \right)   \nonumber \\
&&+\left( D_{G}-d_{G}\right) \left( DC\right) \left( D\psi \right) \left( D%
\overline{D}\psi \right) {\LARGE ]}  \nonumber \\
&&+\frac{C_{V}.D_{G}}{16\pi }\left( 1-\alpha ^{2}\right) \int
d^{2}zd^{2}\theta \epsilon \left( x\right) {\LARGE [}C[\partial Tr\left(
\overline{D}U_{0}DU_{0}^{-1}\right) -\partial \psi Tr\left( \overline{D}%
U_{0}DU_{0}^{-1}\right) ]  \nonumber \\
&&-\frac{1}{2}\left( DC\right) [DTr\left( \overline{D}U_{0}DU_{0}^{-1}%
\right) -D\psi Tr\left( \overline{D}U_{0}DU_{0}^{-1}\right) ]{\LARGE ]},
\end{eqnarray}

\bigskip

where the constant $C_{V}$ is related to the dual Coxeter number of the
superalgebra and is given by:

\bigskip 

\begin{equation}
f_{AB}^{C}f_{AB}^{D}\delta _{CD}+g_{ab}^{A}g_{ab}^{B}\delta
_{AB}=C_{V}.D_{G}. 
\end{equation}

\bigskip

As usual, the Ward-Takahachi identity is obtained by equating the variation
of the action with the variation of the measure. If we dodge the intricate
details of the well known identities \cite{fujikawa} which are pertinent to
our cumbersome calculation, we obtain, finally, the basic super B.R.S
identity:

\bigskip

\begin{eqnarray}
\overline{D}J_{\theta } &=&\overline{D}{\LARGE \langle }\left(
D_{G}-d_{G}-10\right) {\LARGE [}C\left( \partial \psi \right) D\psi +\frac{1%
}{2}C\left( \partial D\psi \right) {\LARGE ]}-4D{\LARGE [}C\partial \psi -%
\frac{1}{2}\left( DC\right) D\psi {\LARGE ]\rangle }  \nonumber \\
&&+\frac{C_{V}.\cdot D_{G}}{8}\left( 1-\alpha ^{2}\right) [\langle
C[\partial Tr\left( \overline{D}U_{0}DU_{0}^{-1}\right) -\partial \psi
Tr\left( \overline{D}U_{0}DU_{0}^{-1}\right) ]  \nonumber \\
&&-\frac{1}{2}\left( DC\right) [DTr\left( \overline{D}U_{0}DU_{0}^{-1}%
\right) -D\psi Tr\left( \overline{D}U_{0}U_{0}^{-1}\right) ]\rangle ]. 
\end{eqnarray}

\bigskip 

This super B.R.S identity exhibits three kinds of anomalies:

(a) It is proportional to $\left( D_{G}-d_{G}-10\right) $\ which determines
the well-known critical dimension for the consistency of the quantum model
based on a super-Lie algebra in a flat space. The dimension $D_{G}-d_{G}$ is
the same as the dimension found by Fujikawa et al. \cite{fujikawa} and the
value $-d_{G}$ is in agreement with the result found by Henningson \cite
{henningson} who studied this model using the covariant quantization method.

(b) It is a total divergence. It is related to the ghost number anomaly, and
exist also in the free case.

\copyright\ It can be eliminated only if $\alpha ^{2}=1$ for supergroups
with $C_{V}\neq 0.$ The addition of a Wess Zumino term to a non linear sigma
model defined on a supergroup manifold leads to a free theory for a certain
value of the relative coupling constant of the two terms in the action $%
\left( \alpha ^{2}=1\right) $. In that case the full action is super
conformally invariant.

\bigskip

\section{Conclusion}

The principles of conformal and B.R.S invariance have deep consequences for
string theory and in quantization of gauge theories. In the present paper,
we have presented the detailed B.R.S analyses of a closed type II
supersymmetric string ( two-dimensional supergravity ) or supersymmetric non
linear sigma-model defined on a group supermanifold built on Lie
superalgebra in the presence of the Wess Zumino term. By a direct evaluation
of the regularization of the super Jacobian corresponding to localised
holomorphic super-BRS transformations of the superstring and its ghosts, we
found that this model is quantum mechanically consistent only in the case where
the relative coupling constant satisfies the well-known relation $\alpha
^{2}=1$ for supergroups with $C_{V}\neq 0$. We are not surprised to find
that this value is the same as the infrared fixed point of the bosonic
sector of the theory \cite{mesref}, the critical point where the chiral Bose
theory with Wess-Zumino term is equivalent to free Fermi theory \cite{witten}%
. Finally, let us emphasise once more that our results are in perfect
agreement with several results based essentially on the computation of the
supersymmetric beta function at one-loop level \cite{abdalla}.

\bigskip

\textbf{Acknowledgments}

The author is very grateful to W. R\"{u}hl and S.V. Ketov for the usual
discussions.

\end{document}